\documentclass[prl,twocolumn,showpacs,amsmath,amssymb]{revtex4}
\usepackage{graphicx}
\usepackage{epstopdf}
\usepackage{enumerate}
\newcommand \beq{\begin{eqnarray}}
\newcommand \eeq{\end{eqnarray}}

\def\simge{\mathrel{%
       \rlap{\raise 0.511ex \hbox{$>$}}{\lower 0.511ex \hbox{$\sim$}}}}
\def\simle{\mathrel{
       \rlap{\raise 0.511ex \hbox{$<$}}{\lower 0.511ex \hbox{$\sim$}}}}
\def\abf{a_{bf}^{\ }}
\def\abfb{a_{bfb}^{\ }}
\def\afbf{a_{fbf}^{\ }}
\def\ab{a_{bb}^{\ }}
\def\af{a_{ff}^{\ }}

\def\gbf{g_{_{bf}}^{\ }}

\def\aN{a_{_{NN}}}

\def\mN{m_{_{N}}}
\def\mb{m_{b}^{\ } }
\def\mf{m_{f}^{\ } }
\def\mR{m_{_{R}} }

\def\mub{\mu_{b}^{\ } }
\def\muf{\mu_{f}^{\ } }

\def\nf{n_{f}^{\ } }

\begin{document}

\title{Simulating Dense QCD Matter with Ultracold Atomic Boson-Fermion Mixtures}
\author{Kenji Maeda,$^{1}$ Gordon Baym,$^{2}$ and Tetsuo Hatsuda$^{1}$  }
\affiliation{
$^{1}$Department of Physics, University of Tokyo, Tokyo 113-0033, Japan \\
$^{2}$Department of Physics, University of Illinois, 1110 W. Green Street,
Urbana, Illinois 61801, USA
}

\begin{abstract}

We delineate, as an analog of two flavor dense quark matter, the phase structure of a many-body mixture of atomic bosons and fermions in two internal states 
with a tunable boson-fermion attraction. The bosons \textit{b} correspond to diquarks, and the fermions \textit{f} to unpaired quarks.  
For weak \textit{b-f} attraction the system is a mixture 
of a Bose-Einstein condensate and degenerate fermions, while for strong attraction composite \textit{b-f} 
fermions \textit{N}, analogs of the nucleon, are formed, which are superfluid due to the \textit{N-N} attraction 
in the spin-singlet channel. 
We determine the symmetry breaking patterns at finite temperature as a function of 
the \textit{b-f} coupling strength, and relate the phase diagram to that of dense QCD.

\end{abstract}
\maketitle

 Ultracold atomic systems and high density QCD matter, although differing by some 20 orders of magnitude in energy scales, share certain analogous physical aspects, e.g., crossovers from Bose-Einstein condensation (BEC) to BCS \cite{regal,Baym:2008me}.   Motivated by phenomenological studies of QCD that indicate a strong spin-singlet diquark correlation inside the nucleon \cite{diquarks}, we focus here on modeling the transition from color superconducting two-flavor quark matter (2SC) at high density to superfluid hadronic matter at low density in terms of a boson-fermion system, in which small size diquarks are the bosons, unpaired quarks the fermions, and the extended nucleons are regarded as composite boson-fermion particles.  Recent advances in atomic physics have made it possible indeed to realize such systems in the laboratory.  In particular, tuning the atomic interaction via a Feshbach resonance allows formation of heteronuclear 
molecules, as recently observed in a mixture of $^{87}$Rb and $^{40}$K atomic vapors in a 3D optical lattice \cite{exp1}, and in an optical dipole trap \cite{exp2}. 
 
 The analogy is incomplete however.
  The gluonic attraction in QCD is a function of the baryon density, and thus 
  tuning the coupling strength at fixed density is not possible in dense matter; furthermore,   
 chiral symmetry breaking plays an important role in the QCD transition \cite{Hatsuda:2006ps}.
 With these reservations in mind, we suggest that fuller understanding, both theoretical and experimental, of the boson-fermion mixture, as well as a mixture of three species of atomic fermions, as discussed in \cite{Rapp:2006rx}, can  reveal properties of high density QCD not readily observable in laboratory experiments.

We first delineate possible phase structures that
can be realized in a mixture of single-component bosons (\textit{b}) and two-component fermions (\textit{f}) 
at finite temperature as a function of the boson-fermion (\textit{b-f}) interaction.
 In weakly coupled \textit{b-f} mixtures, an induced interaction 
 between bosons arising from  density fluctuations of fermons 
 modifies the critical temperature of the Bose condensate.  In addition,
 an induced interaction between fermions due to density fluctuations of bosons 
 may lead to superfluidity of the fermions \cite{bf2}. 
 On the other hand, in strongly coupled mixtures, boson-fermion molecules 
 are formed \cite{bf4}, which may become  superfluid \cite{bf5}.  By  analyzing
  the realization of internal symmetries in each phase, we classify the types of phase 
  boundaries.  We then discuss the detailed connection with the hadronization phase transition
  in QCD matter at low temperatures.  
 
 We consider a (nonrelativistic) boson-fermion mixture with Hamiltonian density, 
 \beq
\mathcal{H} \!&=&\! \frac1{2\mb}|\nabla\phi(x)|^2 - \mub
|\phi(x)|^2 + \frac12\bar{g}_{_{bb}}|\phi(x)|^4 \nonumber\\
    \!&+&\! 
       \sum_{\sigma}\biggl(\frac1{2\mf}|\nabla\psi_{\sigma}(x)|^2-\muf
       |\psi_{\sigma}(x)|^2  
       \biggl) \nonumber\\
    \!&+&\!  
       \bar{g}_{_{ff}}|\psi _{\uparrow}(x)|^2  |\psi_{\downarrow}(x)|^2  
         +\sum_{\sigma}\bar{g}_{_{bf}}|\phi(x)|^2 |\psi_{\sigma}(x)|^2 , 
\nonumber \\
\label{eq:bf-hamiltonian}
\eeq
where $\phi$ is the boson and $\psi$ the fermion  field. 
 We label the two internal states of the fermions by spin indices 
 $\sigma =\{\uparrow ,\downarrow \}$. 
 We focus for simplicity on an equally populated mixture of $n$ bosons and $n$
fermions with 
$n_{\uparrow}= n_{\downarrow}=n/2$.  

The bare boson-fermion coupling $\bar{g}_{_{bf}}$ is related to the 
 renormalized coupling $\gbf$ and to
the \textit{s} wave scattering length $\abf$ by
\beq
\frac{\mR}{2\pi \abf} 
=\frac{1}{g_{_{bf}}} 
=\frac{1}{\bar{g}_{_{bf}}}+
 \int_{|{\bf k}| \le \Lambda}\frac{d^3k}{(2\pi)^3} 
 \frac{1}{\varepsilon_{b}(k)+ \varepsilon_{f}(k)}\: ,  
\eeq 
where $\varepsilon_{i}(k)= k^2/2m_i$ ($i=b, f$) is the single-particle kinetic energy,
$\mR$ is the boson-fermion reduced mass, and $\Lambda$ is a high momentum cutoff. We define
 $r_0 \equiv (2\Lambda/\pi)^{-1}$ as a  
typical atomic scale.
We assume an attractive bare \textit{b-f} interaction  ($\bar{g}_{_{bf}}\!\!<\! 0 $), tunable in magnitude, 
 with $\Lambda $ fixed so that the scattering length $\abf$ can change sign: 
 namely, $\abf \rightarrow \bar{g}_{_{bf}} \mR/2\pi$ for small negative
$\bar{g}_{_{bf}}$, while $\abf \rightarrow r_0$ for large negative
$\bar{g}_{_{bf}}$.
We keep the bare boson-boson and fermion-fermion interactions
fixed and repulsive ($\bar{g}_{_{bb}}\! > \! 0, ~\bar{g}_{_{ff}} \!\!>\! 0$); thus, 
the renormalized couplings 
$g_{_{bb}}=4\pi a_{bb}/\mb$ and $g_{_{ff}}=4\pi a_{ff}/\mf$ 
are always positive. 
These repulsions are required for the stability of the bosons at weak bare \textit{b-f} coupling 
and also for the dominance of  two-body molecules (\textit{bf}) 
over three-body molecules (\textit{bbf}) and (\textit{bff}) at strong bare \textit{b-f} attraction \cite{threebody}. 
Boson-fermion mixtures with an attractive \textit{b-f} interaction 
%($\abf<0$) 
may be realized in three-component cold atomic experiments using, e.g., the hyperfine state 
 $|f=1,m_f=1 \rangle$ 
 of $^{87}$Rb, mixed with the hyperfine states 
 $|9/2,-5/2 \rangle $ and $|9/2,-9/2 \rangle$ of
 $^{40}$K \cite{bb:ff:bf:expFesh}.

%%% FIG.1 U_{bfb} %%%
\begin{figure}[b]
\begin{center}
\includegraphics[width=4.0cm]{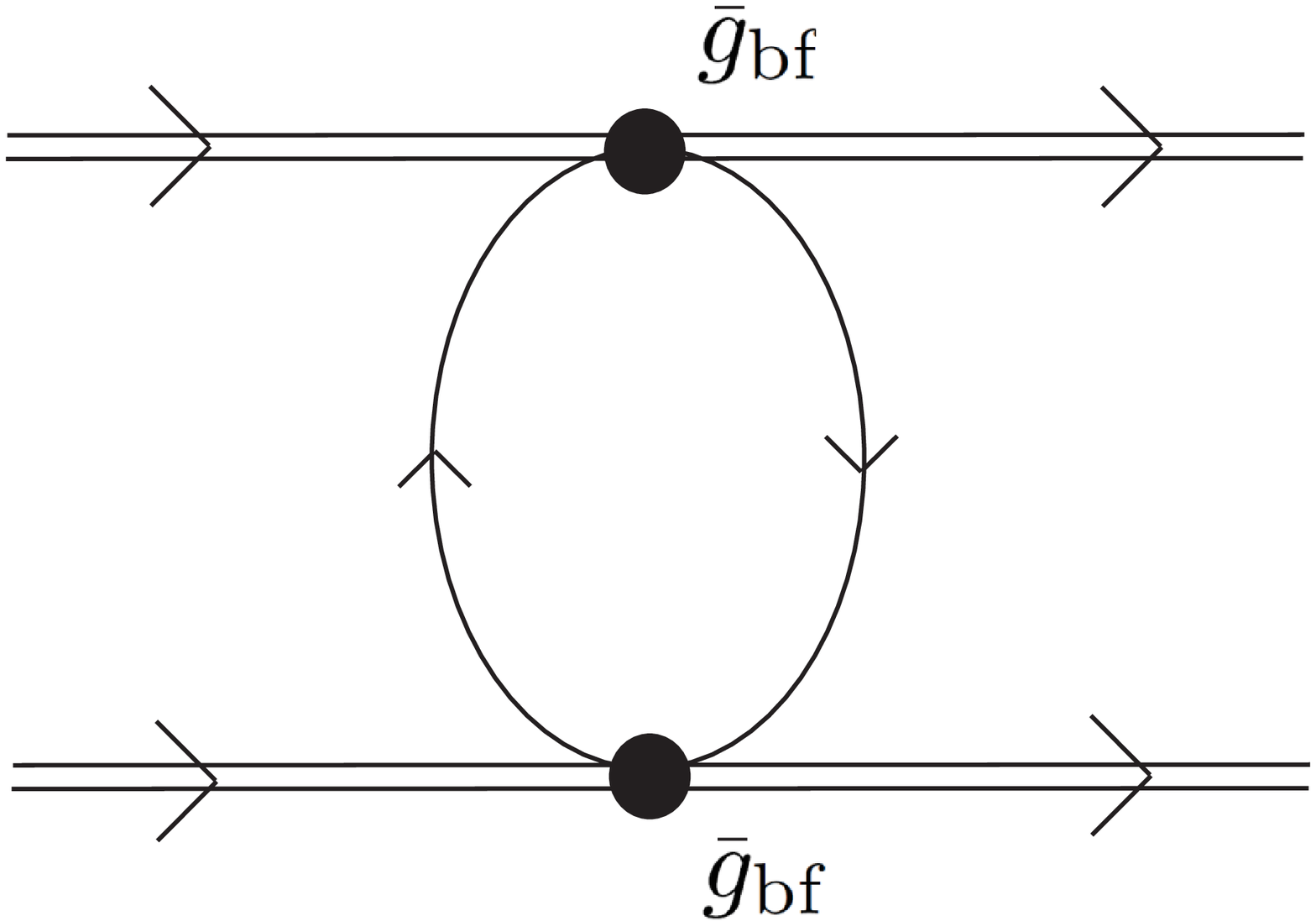}
\end{center}
\vspace{-0.5cm}
\caption{Fermion-induced interaction between bosons
to second order in the boson-fermion interaction.
The single lines denote free fermions and the double lines free bosons.
}
\label{Fig1}
\end{figure}

We first lay out the phase structure at weak bare \textit{b-f} coupling, where the dimensionless  parameter 
$\eta\equiv  -1/n^{1/3}\abf$ is $\gg 1$. 
In the absence of  \textit{b-f} attraction ($\eta = + \infty$) with weak \textit{b-b} repulsion 
($n^{1/3}\ab \ll 1 $), 
boson condensation (\textit{b}-BEC)  occurs below the critical temperature,
$T^0 _{c}(b\mathchar`- \mathrm{BEC})
\simeq (1+1.32n^{1/3}\ab+\cdots)T_0$, with the ideal BEC transition temperature
 $T_0\equiv 2\pi[n/\zeta(3/2)]^{2/3}/\mb$ \cite{arnold02}.
 A weak \textit{b-f} interaction induces an attraction between bosons via fermion  density fluctuations (Fig.~\ref{Fig1}) given by 
 $U_{bfb}(p,\omega;T)
=-\bar{g}_{_{bf}}  ^2 \Pi(p,\omega;T)$, where $\Pi(p,\omega;T)$ is the 
one-loop fermionic polarization at temperature $T$.  In the static limit, relevant for elastic scattering of bosons, 
\beq 
\Pi(p,0;T)=-2\!\int\!\frac{d^3 q}{(2\pi)^3}
\frac{\nf(|\bold{p+q}|)-\nf(q)}{\varepsilon_{f}(|\bold{p+q}|)-\varepsilon_{f}(q)},\:
\eeq
with
$\nf(p)=1/[e^{[\varepsilon_{f}(p)-\muf]/T}+1]$ the Fermi occupation.
 Then, the characteristic length of the fermion-induced
 \textit{b-b} interaction near $T_{c}(b\mathchar`- \mathrm{BEC})$ is 
$
\abfb \equiv (\mb/4\pi)
U_{bfb}(\sqrt{2\mb T_0}, 0 ; T_0) < 0 
$,
where we set $T=T_0$ and take a typical momentum transfer $p$ equal to 
the thermal momentum $\sqrt{2\mb T_0}$. 
For  $\mb/\mf = 2$, corresponding to a $^{87}$Rb-$^{40}$K mixture, 
$k_{_{F}} \abfb \simeq -11\:\eta^{-2}$ with 
$k_{_{F}} =(3\pi^2 n)^{1/3}$. 
Since the induced attraction tames the bare \textit{b-b}  repulsion, 
the transition temperature for \textit{b}-BEC 
decreases to 
\beq
T_{c}(b\mathchar`- \mathrm{BEC})\:=\:
[1+1.32n^{1/3}(\ab+\abfb)+\cdots]\:T_0 \:.    
\label{eq:Tcb}
\eeq
The replacement $\ab \rightarrow \ab + \abfb$
is exact to second order in the \textit{b-f} interaction;
more generally such a replacement is exact in a large ${\cal N}$ extension ($\abf, \ab \sim 1/{\cal N}$)
where only bubble summations are important \cite{BBZ,LargeN}.
Calculation of higher order corrections to 
$T_{c}(b\mathchar`- \mathrm{BEC})$ is
beyond our present scope.
The Bose gas becomes mechanically unstable 
when the net \textit{b-b} attraction becomes large enough to overcome the thermal pressure \cite{baym-mueller}:
$3nT/2+(g_{_{bb}}+U_{bfb})n^2/2 <0$. 
For $\mb/\mf = 2$ and $n^{1/3}\ab =10^{-2}$, the critical values for the instability are  
$\eta_c ^{b\mathchar`- {\rm BEC}} \simeq 21$ at $T=0$ and 
$\eta_c ^{ b\mathchar`- {\rm BEC} }\simeq 2.2$ 
near $T _c (b\mathchar`- {\rm BEC})$.  At high $T$, thermal pressure stabilizes the system.

A weak \textit{b-f} interaction  also leads to a boson-induced attraction between fermions
similar to the phonon-induced attraction between electrons in metals \cite{bf2}.
The characteristic length of the induced \textit{f-f}  interaction in fully condensed bosons, 
$\afbf$, averaged over the Fermi surface, is 
$
\afbf \equiv
 (\pi/3)^{1/3}(\mb\mf/\mR^2)\ln (1+x^2)(k_{_{F}}\eta ^2)^{-1}
%\label{eq:afbf}
$,
with $x=\sqrt{3\pi/4k_{_{F}}\ab}$.  
 For $\mb/\mf = 2$, 
$k_{_{F}}a_{fbf}\simeq -4.6\ln (1+x^2) \eta^{-2}$. 
A net attraction, $\af + \afbf < 0$, causes fermionic BCS superfluidity (\textit{f}-BCS) 
below a critical temperature \cite{gmb61}
\beq 
T_c(f\mathchar`- {\rm BCS})
= \frac{e^\gamma}{\pi}\biggl( \frac{2}{e}\biggl)^{7/3} \varepsilon _{_{F}}
  \exp \left( \frac{\pi}{2 k_{_{F}}(\af + \afbf )}  \right) ,
\label{eq:Tcf}
\eeq
 where $\varepsilon_{_{F}}=k_{_{F}} ^2/2\mf$ and 
 $\gamma$ is Euler's constant.
For $\mb /\mf=2$, \textit{f}-BCS emerges 
when $\eta<\eta_c ^{f\mathchar`- {\rm BCS}}
= 2.1[\ln (1+x^2)/k_{_{F}} \af ]^{1/2}$. 
As the \textit{b-f} attraction increases, 
$T_{c}(f\mathchar`- \mathrm{BCS})$  increases due to the increase of the net \textit{f-f}  attraction. 
%FIG.2
\begin{figure}[b]
\begin{center}
\includegraphics[width=5.5cm]{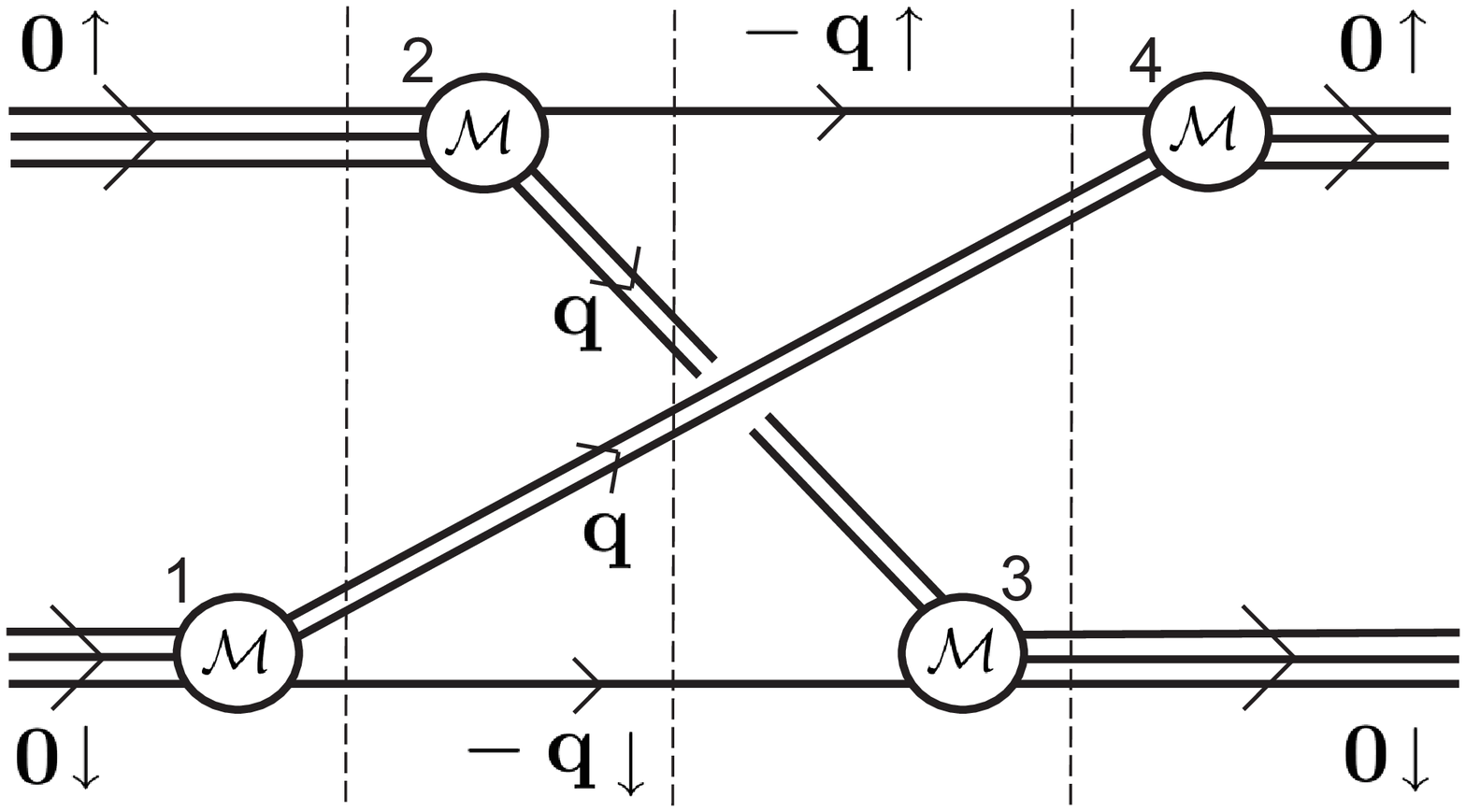}
\end{center}
\vspace{-0.5cm}
\caption{$T$-matrix diagram for scattering of composite fermions of opposite spins, arising from
exchange of the constituent bosons.
The single, double, and triple lines denote free
 fermions, bosons and composite fermions, respectively, and the three dashed lines indicate the intermediate states.
}
\label{Fig2}
\end{figure}

 Let us turn to the regime of strong bare \textit{b-f} coupling where $\eta$ is large and negative.
 Here bound molecules or {\it composite fermions}, \textit{N} = (\textit{bf}), are formed 
with a kinetic mass $\mN=\mb+\mf$. 
We estimate the \textit{s} wave scattering length of two \textit{N}'s of opposite spins from the  diagram sketched in Fig.~\ref{Fig2}.
% Spin indices of the composite fermions are 
%those of their constituent fermions. 
Integration over the three intermediate states for this process yields
the corresponding  $T$-matrix element  
with  zero initial and final momenta, 
\beq 
T_{NN}(\bold{0},\bold{0})=
4\int_{|{\bf q}| \le \Lambda} \frac{d^3 q}{(2\pi)^3}
\frac{|\mathcal{M}|^4}{2[\epsilon_{N}- \varepsilon_{b}(q)-\varepsilon_{f}(q)]^3},
\label{eq:tmatrix}
\eeq
where $\epsilon_{N}$ is the binding energy of the composite fermion. 
The prefactor 4 arises from four possible time orderings of 
the dissociation and recombination,
(12;34) as in Fig.~\ref{Fig2}, (21;34), (12;43), and (21;43).  Using 
 the Schr\"{o}dinger equation,
$[-\nabla ^2/2\mR+V(\bold{r})]\psi_{N}(\bold{r})
=\epsilon_{N}\psi_{N}(\bold{r})$, where $V(\bold{r})$ is the bare \textit{b-f} potential,
$\psi_{N}(\bold{r})$ the internal wave function of the composite fermion, 
and $\psi_{N}(\bold{q})$ its Fourier transform,
 we write the matrix element $\mathcal{M}$ for dissociation of \textit{N} into \textit{b} and \textit{f} as 
$
\mathcal{M} = (\epsilon_{N}-q^2/2\mR)\psi_{N}(\bold{q});
$
thus, 
\beq
T_{NN}(\bold{0},\bold{0})=
2\int_{|{\bf q}| \le \Lambda} \frac{d^3 q}{(2\pi)^3}|\psi_N (q)|^4(
\epsilon_{N}- q^2/2\mR).
\label{eq:tmatrix1}
\eeq

For a bound state in a zero-range potential, $\psi_N(\bold{r})= e^{-r/\abf}/ r\sqrt{2\pi \abf }$ 
and $\epsilon_{N}= -1/2\mR a^2_{bf}$, and thus
$\mathcal{M}=-\sqrt{2\pi/\mR^2\abf}$ \cite{notice}.   Equation~(\ref{eq:tmatrix1}) yields 
the \textit{s} wave scattering length for composite fermions of opposite spins
 for negative  and large $\eta$ 
\beq
\aN = \frac{\mN}{4\pi}T_{NN} (\bold{0},\bold{0})
=- \frac{\mN}{2\mR} \abf\Gamma . 
\label{eq:anborn}
\eeq
Here $0.76 < \Gamma(\abf/r_0) <1 $ in the interval $1 < \abf/r_0 < \infty$.
This result is the same in magnitude but opposite in sign from the scattering length between
difermion molecules in the same approximation \cite{bf5}. 
For $\mb/\mf=2$, we obtain $k_{_{F}} \aN \simeq -7.0 \Gamma/|\eta|$.
 The above estimate of the \textit{N-N} scattering length is the leading order term in 
  the large ${\cal N}$ extension of the present boson-fermion model \cite{LargeN}.
 For finite ${\cal N}$, it is corrected by
 (i) other recombination diagrams  \cite{Petrov}, (ii) short range \textit{b-b}  and \textit{f-f}  repulsions,
and (iii) possible contributions of trimers such as (\textit{bff}) and (\textit{bbf}) \cite{threebody}.   
We leave the  calculation of such  corrections for the future 
and treat Eq.~(\ref{eq:anborn}) as a first estimate.
Equation~(\ref{eq:anborn}) implies that the low energy effective interaction
between composite fermions in the spin-singlet channel is weakly attractive; the stronger 
the bare \textit{b-f} coupling,
the weaker the \textit{N-N} interaction. 
Such an effective attraction causes composite fermions 
to become BCS-paired (\textit{N}-BCS) below a transition temperature,
\beq 
T_c(N\mathchar`- \mathrm{BCS})
&=& \frac{e^\gamma}{\pi}\biggl( \frac{2}{e}\biggl)^{7/3} \varepsilon_{_N}
  e^{\pi/2 k_{_{F}} \aN} \: .
  \label{eq:TcN}
\eeq
where $\varepsilon_{_N}=k_{_{F}} ^2/2\mN$ is the  Fermi energy of the \textit{N}.

\begin{figure}[t]
\begin{center}
\includegraphics[width=7cm]{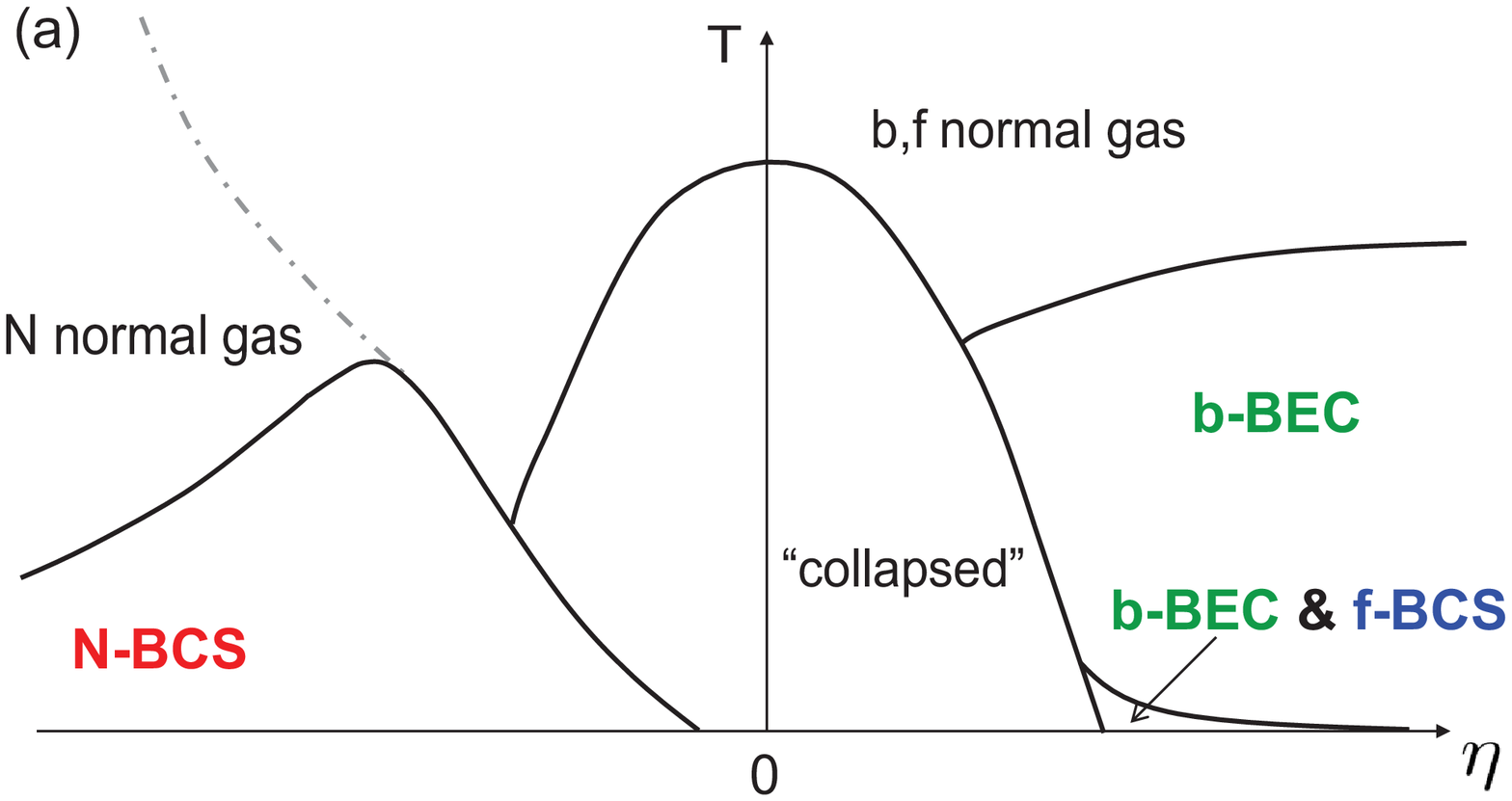}
\
\includegraphics[width=7cm]{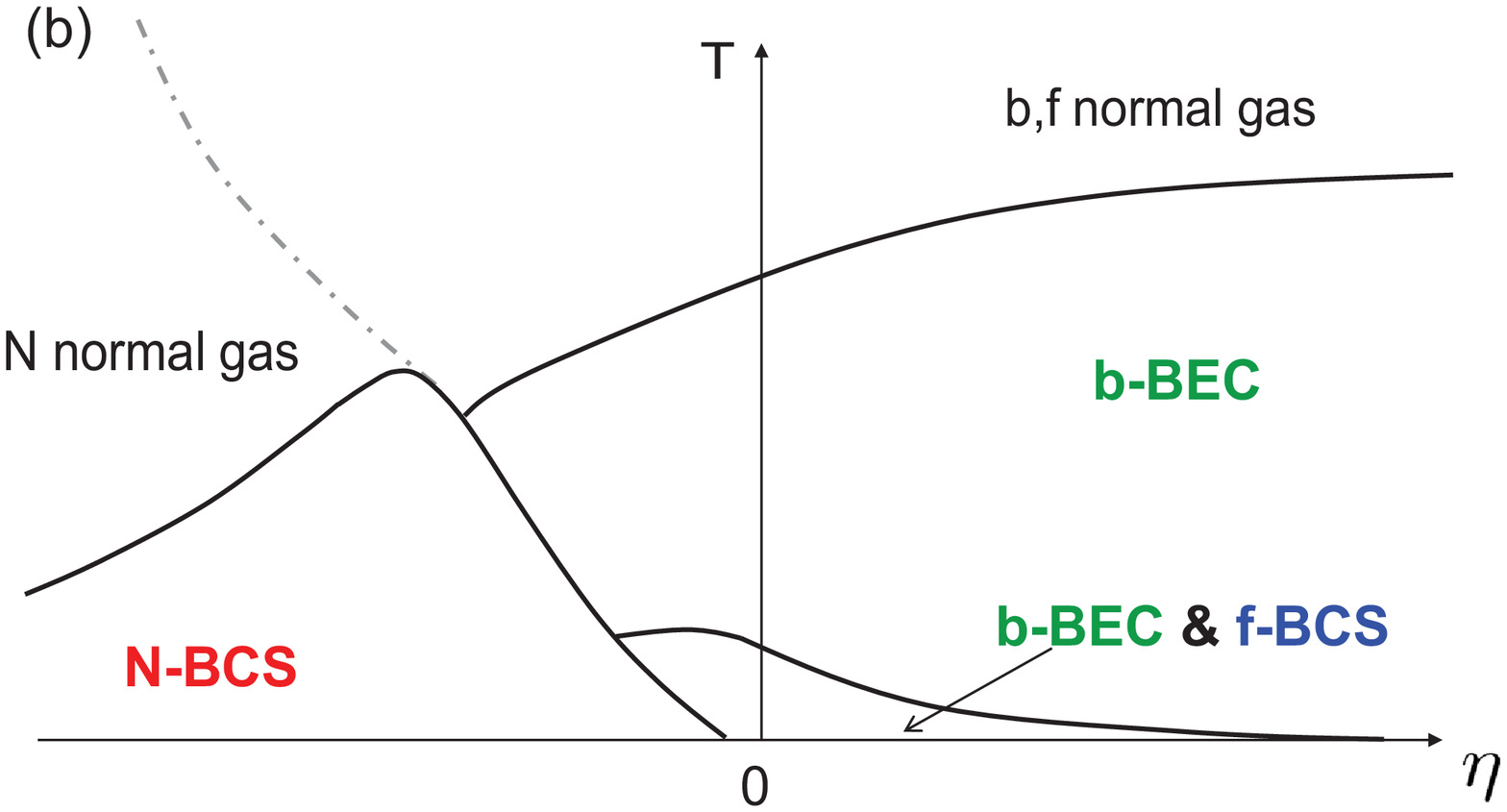}
\end{center}
\vspace{-0.5cm}
\caption{Schematic phase structures of boson-fermion mixtures
with a tunable boson-fermion interaction. 
(a) Weak ${\bar g}_{_{bb}}$; in the regime labeled ``collapsed" the effective boson-boson interaction is negative.
%region, an unstable (collapsed) region 
%of uniform \textit{b}-BEC. 
(b) Strong ${\bar g}_{_{bb}}$; a possibly ``collapsed" region is not shown.}
\label{Fig3}
\end{figure}

 The above analyses suggest possible phase structures of boson-fermion mixtures 
in the $T\mathchar`-\eta$ plane, shown in Fig.~\ref{Fig3}. 
The system at low temperature for $\eta = \infty$ is a mixture of 
the \textit{b}-BEC and degenerate unpaired fermions.
As $\eta$ decreases from the right, the size of the \textit{b}-BEC region shrinks, Eq.~(\ref{eq:Tcb}).  
The \textit{f}-BCS state emerges at $T=T_c$(\textit{f}-BCS), Eq.~(\ref{eq:Tcf}). 
Both effects are caused by the induced attractions.   
On the other hand, for strong bare \textit{b-f} coupling ($\eta$ large and negative), 
the system at low temperature is
in the \textit{N}-BCS state. As $\eta$ increases from the left, the size of \textit{N}-BCS region increases according to Eq.~(\ref{eq:TcN}).
 
At intermediate bare \textit{b-f} coupling ($\eta \sim 0$) where a transition 
from the \textit{b}-BEC (and at low $T$ with coexisting \textit{f}-BCS) phase to \textit{N}-BCS takes place, 
the phase diagram has complex structures 
depending on the relative magnitudes of $\bar{g}_{_{bb}}$, $\bar{g}_{_{ff}}$, and 
$\bar{g}_{_{bf}}$. 
When the intrinsic \textit{b-b}  repulsion is weak, 
uniform \textit{b}-BEC at low temperature becomes unstable 
at $\eta = \eta_c ^{b\mathchar`- {\rm BEC} } > 0$ due to the net \textit{b-b}  attraction, 
while at high temperature the system becomes a stable normal gas of bosons and fermions 
because of thermal pressure \cite{baym-mueller}. 
This situation is indicated by the ``collapsed" region at $\eta \sim 0$ in Fig.~\ref{Fig3}(a). 
For strong bare \textit{b-b}  repulsion, 
the weak coupling formula (\ref{eq:Tcb}) is no longer valid,  
and the \textit{b}-BEC phase may possibly survive into the region $\eta < 0$ 
until the bosons become bound in composite fermions as shown in Fig.~3(b).   To determine the physical structure of the ``collapsed" region requires evaluating the free energy to higher order.

%%% FIG.3 Phase structure %%%

To identify the precise phase boundaries in the region
 $\eta \sim 0$,  we need to solve the system beyond the weak or strong coupling
 regimes studied here. 
%Possible role of the Efimov trimers such as (bbf) and (bff)
% at $\eta \sim 0$ should be also considered \cite{threebody}.
Rather, we classify the phases of the boson-fermion mixture 
by their realizations of the internal symmetry of the system.
We focus only on the continuous symmetries here.
The Hamiltonian density, Eq.~(\ref{eq:bf-hamiltonian}), 
has $U(1)_b \times U(1)_{f _{\uparrow}} \times U(1)_{f _{\downarrow}}$ 
symmetry corresponding to independent  phase rotations of 
$\phi$, $\psi_{\uparrow}$, and $\psi_{\downarrow}$. 
On the other hand, \textit{b}-BEC, \textit{f}-BCS and \textit{N}-BCS break 
$U(1)_b$, $U(1)_{f _{\uparrow}+f _{\downarrow}}$  
and $U(1)_{b+(f _{\uparrow}+f _{\downarrow})}$ symmetries, respectively. 
Here $U(1)_{A \pm B}$ denotes an in-phase rotation of $A$ and $B$ 
for ($+$), and an opposite-phase rotation for ($-$). 
Therefore, in the coexisting  \textit{b}-BEC and \textit{f}-BCS phase, the 
symmetry breaking pattern is  
$U(1)_b \times U(1)_{f _{\uparrow}}\times U(1)_{f _{\downarrow}}
 \rightarrow U(1)_{f _{\uparrow}-f _{\downarrow}}$. On the other hand, 
the \textit{N}-BCS phase has the symmetry breaking pattern, 
$U(1)_b \times U(1)_{f _{\uparrow}}\times U(1)_{f _{\downarrow}} 
\rightarrow U(1)_{b-(f _{\uparrow}+f _{\downarrow})}\times U(1)_{f _{\uparrow}-f _{\downarrow}}$. 
The difference of unbroken symmetries between these two phases 
implies the existence of a well-defined phase boundary, as indicated in Fig.~\ref{Fig3}b, 
 in contrast to the continuous crossover from BEC to BCS in a two-component Fermi system 
\cite{BEC_BCS2}.

Interesting  problems remaining for further research on the phase structure of the mixtures 
include understanding the \textit{N-N}scattering length  beyond  leading order in large ${\cal N}$, 
the quantitative description of the phase structure in the intermediate  coupling regime, 
and extensions to spin-dependent \textit{b-f} interactions  and unequal population of bosons and fermions.

The phase structures we find for boson-fermion mixtures of cold atoms display features of those in two-flavor QCD 
with equal numbers of up (u) and down (d) quarks with three colors (R,G,B). 
The ground state of this system at high density is a 2-flavor color superconductivity (2SC) 
with \textit{s} wave spin-singlet pairing, e.g., between uR and dG, in the color antisymmetric and flavor antisymmetric channel. 
The order parameter for color-symmetry breaking, 
$SU(3)_c \rightarrow SU(2)_c$, is the diquark condensate $ \langle b_3 \rangle$
 with the diquark operator $b_{\gamma}=  \epsilon_{ij} 
\epsilon_{\alpha \beta \gamma} q_{\alpha}^i C \gamma_5 q_{\beta}^j$; here $i,j$ are
 flavor and $\alpha, \beta$ color indices, and 
 $C$ denotes charge conjugation. The  
gap is of order a few tens of MeV;
the remaining quarks, uB and dB, are unpaired 
and form degenerate Fermi seas \cite{alf08}.  On the other hand, 
the ground state of two-flavor QCD with equal numbers of u and d quarks at low density
is nuclear matter with equal numbers of neutrons and protons, a superfluid state
with a pairing gap of a few MeV (see, e.g., \cite{Dean:2002zx}); 
the order parameter for the spontaneous breaking of 
 baryon-number symmetry $U(1)_B$ is the six-quark condensate 
$ \langle N^i_{\uparrow} N^j_{\downarrow} \rangle 
= \langle (b_{\alpha}^{ }q_{\alpha,\uparrow}^i)(b_{\beta}^{ }q_{\beta,\downarrow}^j) \rangle $.

If we model the nucleon, of radius $r_{_N} \sim 0.86 $ fm,
 as a bound molecule of a diquark (of radius $r_{_{D}} \sim 0.5$ fm) and 
 an unpaired quark, we can make the following correspondence between
  cold atoms and QCD:
  \textit{b} $\leftrightarrow$ 2SC-diquarks and \textit{f} $\leftrightarrow$ unpaired-quark,
 \textit{N} $\leftrightarrow$ nucleon, \textit{b-f} attraction $\leftrightarrow$ 
 gluonic attraction, \textit{b}-BEC $\leftrightarrow$ 2SC, and 
 \textit{N}-BCS $\leftrightarrow$ nucleon superfluidity 
 \cite{QCDfBCS}.  In particular, the boundary between the \textit{b}-BEC and \textit{N}-BCS phases shown in 
 Fig.~3(b) is strongly suggestive of that deduced between the color superconducting quark phase and the superfluid hadronic phase in Refs.~\cite{Hatsuda:2006ps,Baym:2008me}.

\begin{acknowledgments}
We thank T. Hirano, Y. Nishida, G. Ripka, S. Sasaki, S. Uchino, and M. Ueda for useful comments.  We are grateful to the Aspen Center for Physics and to the ECT* in Trento,  where parts of this work were carried out.  This research was supported in part by NSF Grant PHY-07-01611 and 
  JSPS Grants-in-Aid for Scientific Research No. 18540253. In addition G.B. wishes to thank the
  G-COE program  of the Physics Department of the University of Tokyo for 
  hospitality and support  during the completion of this work. 
\end{acknowledgments}

%%% REFERENCES %%%


\vspace{-12pt}
\begin{thebibliography}{99}
\vspace{-12pt}

\bibitem{regal} 
C.~A.~Regal, M.~Greiner, and D.~S.~Jin, Phys. Rev. Lett. {\bf 92}, 040403 (2004).

\bibitem{Baym:2008me}
  G.~Baym,  T.~Hatsuda, M.~Tachibana and N.~Yamamoto,
  %``The axial anomaly and the phases of dense QCD,''
  J.\ Phys.\ G {\bf 35}, 104021 (2008).

\bibitem{diquarks}
M.~Anselmino et al., 
%E.~Predazzi, S.~Ekelin, S.~Fredriksson, and D.~B.~Lichtenberg, 
%``Diquarks,'' 
Rev.\ Mod.\ Phys.\ {\bf 65}, 1199 (1993); 
R.~L.~Jaffe, 
%``Exotica,''
Phys.\ Rept.\ {\bf 409}, 1 (2005); 
A.~Selem and F.~Wilczek, 
%in``Hadron systematics and emergent diquarks,'' 
%\textit {Ringberg 2005, New trends in HERA physics}
arXiv:hep-ph/0602128.

%\bibitem{Feshbach} Reviewed in 
%I.~Bloch, J.~Dalibard and
% W.~Zwerger, Rev. Mod. Phys. {\bf 80}, 885 (2008).

\bibitem{exp1} C.~Ospelkaus et al., Phys. Rev. Lett. {\bf 97}, 120402 (2006).
 
\bibitem{exp2}
J.~J.~Zirbel et al.,  Phys. Rev. A {\bf 78}, 013416 (2008).

 
\bibitem{Hatsuda:2006ps}
  T.~Hatsuda, M.~Tachibana, N.~Yamamoto, and G.~Baym,
  %``New critical point induced by the axial anomaly in dense QCD,''
  Phys.\ Rev.\ Lett.\  {\bf 97}, 122001 (2006).

\bibitem{Rapp:2006rx}
A.~Rapp, G.~Zarand, C.~Honerkamp, and W.~Hofstetter, 
%``Color superfluidity and 'baryon' formation in ultracold fermions,'' 
Phys.\ Rev.\ Lett.\ {\bf 98}, 160405 (2007);
R.~W.~Cherng, G.~Refael, and E.~Demler,
Phys.\ Rev.\ Lett.\ {\bf 99}, 130406 (2007).

 
\bibitem{bf2} M.~J.~Bijlsma, B.~A.~Heringa, and H.T.C.~Stoof,
Phys. Rev. A {\bf 61}, 053601 (2000); 
H.~Heiselberg, C.~J.~Pethick, H.~Smith, and L.~Viverit, 
Phys. Rev. Lett. {\bf 85}, 2418 (2000).
 
\bibitem{bf4} A.~Storozhenko et al., 
%P.~Schuck, T.~Suzuki, H.~Yabu, and J.~Dukelsky, 
 Phys. Rev. A {\bf 71}, 063617 (2005);
T.~Watanabe, T.~Suzuki, and P.~Schuck, Phys. Rev. A {\bf 78}, 033601 (2008).

\bibitem{bf5} M.~Yu.~Kagan,
 I.~V.~Brodsky, D.~V.~Efremov, and A.~V.~Klaptsov,
Phys. Rev. A {\bf 70}, 023607 (2004).

\bibitem{threebody}
In  a mixture with equal numbers of  \textit{b} and \textit{f} 
with \textit{b-b}  and \textit{f-f}  repulsion, as here, the state 
(\textit{bf})+(\textit{bf}) is energetically more favorable than (\textit{bbf})+\textit{f} and (\textit{bff})+\textit{b} 
for $\abf \rightarrow r_0$.  For $| \abf | \to \infty$, 
(\textit{bbf}) and (\textit{bff}) as Efimov trimers may play important roles although 
they do not appear as bound states in the large ${\cal N}$ model discussed below.
      
\bibitem{bb:ff:bf:expFesh}
See, e.g., C.~Klempt et al., Phys. Rev. A {\bf 76}, 020701(R) (2007); 
A.~Simoni et al.,  Phys. Rev. A {\bf 77}, 052705 (2008).

\bibitem{arnold02} 
G.~Baym et al., 
%J.-P.~Blaizot, M.~Holzmann, F.~Lalo\"e, and D.~Vautherin, 
Phys. Rev. Lett {\bf 83}, 1703 (1999); 
P.~Arnold, G.~Moore, and B.~Tom\'{a}\v{s}ik,
Phys. Rev. A {\bf 65}, 013606 (2001). 

\bibitem{BBZ}
G.~Baym, J.-P.~Blaizot, and J.~Zinn-Justin, 
Europhys. Lett. {\bf 49}, 150 (2000).

\bibitem{LargeN} 
The interaction in the extension of Eq.~(\ref{eq:bf-hamiltonian}) 
to large ${\cal N}$ is  
$(1/{\cal N})\sum_{i,j}^{\cal N}[
\frac12\bar{g}_{_{bb}}(\phi^{\dagger}_i\phi^{\dagger}_i)(\phi_j\phi_j)
+\bar{g}_{_{ff}}(\psi^{\dagger}_{\uparrow i} \psi^{\dagger}_{\downarrow i})
(\psi_{\downarrow j}\psi_{\uparrow j})$ 
$+\bar{g}_{_{bf}}\sum_\sigma(\phi^{\dagger}_i \psi^{\dagger}_{\sigma i})
(\phi_j \psi_{\sigma j})]. $


\bibitem{baym-mueller}
E.~J.~Mueller and G.~Baym, Phys. Rev. A {\bf 62}, 053605 (2000).

\bibitem{gmb61} L.~P. Gorkov and T.~K.~Melik-Barkhudarov, 
Sov. Phys. JETP {\bf 13}, 1018 (1961).
  
\bibitem{notice} 
For $\abf$ close to $r_0$, estimating $\mathcal{M}$ requires using a
 wave function obtained from a more realistic atomic potential.
 
\bibitem{Petrov}
D.~S.~Petrov, C.~Salomon and G.~V.~Shlyapnikov, Phys. Rev. A {\bf 71}, 012708 (2005);
I.~V.~Brodsky et al., 
%M.~Y.~Kagan, A.~V.~Klaptsov, R.~Combescot, and X.~Leyronas, 
Phys. Rev. A {\bf 73}, 032724 (2006);
J.~Levinsen and V.~Gurarie, Phys. Rev. A {\bf 73}, 053607 (2006).

%\bibitem{yabu}
%T.~Nishimura, A.~Matsumoto, and H.~Yabu,
%Phys. Rev. A {\bf 77}, 063612 (2008).

\bibitem{BEC_BCS2} A.~J.~Leggett, in \textit{Modern Trends in the Theory of
Condensed Matter},  edited by A.~Pekalski and R.~Przystawa 
(Springer-Verlag, Berlin, 1980); 
P.~Nozi\`{e}res and S.~Schmitt-Rink,
J. Low. Temp. Phys. {\bf 59}, 195 (1985).

\bibitem{alf08} M.~G.~Alford, A.~Schmitt, K.~Rajagopal, and T.~Sch\"afer, 
Rev. Mod. Phys. {\bf 80}, 1455 (2008).

\bibitem{Dean:2002zx}
D.~J.~Dean and M.~Hjorth-Jensen, 
%``Pairing in nuclear systems: from neutron stars to finite nuclei,'' 
Rev.\ Mod.\ Phys.\ {\bf 75}, 607 (2003).


\bibitem{QCDfBCS}
 The \textit{s} wave, spin singlet and color symmetric pairing
 corresponding to \textit{f}-BCS is unlikely in QCD because of the repulsive
 one-gluon-exchange in this channel.

\end{thebibliography}
\end{document}